\documentclass[pdflatex,sn-mathphys-num]{sn-jnl}
\usepackage{geometry}
\usepackage{graphicx}%
\usepackage{multirow}%
\usepackage{amsmath,amssymb,amsfonts}%
\usepackage{amsthm}%
\usepackage{bm}%
\usepackage{mathrsfs}%
\usepackage[title]{appendix}%
\usepackage{xcolor}%
\usepackage{textcomp}%
\usepackage{manyfoot}%
\usepackage{booktabs}%
\usepackage{algorithm}%
\usepackage{algorithmicx}%
\usepackage{algpseudocode}%
\usepackage{listings}%

\theoremstyle{thmstyleone}%
\theoremstyle{thmstyletwo}%

\theoremstyle{thmstylethree}%

\begin{document}

\title[Article Title]{Realizing the Haldane Model in Thermal Atoms}

\author[1,2]{\fnm{Jiefei} \sur{Wang}}
\equalcont{These authors contributed equally to this work.}

\author[1]{\fnm{Jianhao} \sur{Dai}}
\equalcont{These authors contributed equally to this work.}

\author[1]{\fnm{Ruosong} \sur{Mao}}
\equalcont{These authors contributed equally to this work.}

\author[1]{\fnm{Yunzhou} \sur{Lu}}

\author[1]{\fnm{Xiao} \sur{Liu}}

\author[2]{\fnm{Huizhu} \sur{Hu}}

\author[1,2,3]{\fnm{Shi-Yao} \sur{Zhu}}

\author*[1]{\fnm{Xingqi} \sur{Xu}}\email{xuxingqi@zju.edu.cn}

\author*[2]{\fnm{Han} \sur{Cai}}\email{hancai@zju.edu.cn}

\author*[1,2,3]{\fnm{Da-Wei} \sur{Wang}}\email{dwwang@zju.edu.cn}

\affil[1]{\orgdiv{Zhejiang Key Laboratory of Micro-Nano Quantum Chips and Quantum Control}, \orgname{School of Physics, and State Key Laboratory for Extreme Photonics and Instrumentation, Zhejiang University}, \orgaddress{\city{Hangzhou}, \postcode{310027}, \state{Zhejiang Province}, \country{China}}}

\affil[2]{\orgdiv{College of Optical Science and Engineering}, \orgname{Zhejiang University}, \orgaddress{\city{Hangzhou}, \postcode{310027}, \state{Zhejiang Province}, \country{China}}}

\affil[3]{\orgdiv{Hefei National Laboratory}, \orgaddress{\city{Hefei}, \postcode{230088}, \state{Anhui Province}, \country{China}}}

\abstract{
Topological materials hold great promise for developing next-generation devices with transport properties that remain resilient in the presence of local imperfections. However, their susceptibility to thermal noise has posed a major challenge. In particular, the Haldane model, a cornerstone in topological physics, generally requires cryogenic temperatures for experimental realization, limiting both the investigation of topologically robust quantum phenomena and their practical applications.
In this work, we demonstrate a room-temperature realization of the Haldane model using atomic ensembles in momentum-space superradiance lattices, a platform intrinsically resistant to thermal noise. 
The topological phase transition is revealed through the superradiant emission contrast between two timed Dicke states in the lattice. Crucially, the thermal resilience of this platform allows us to access a deep modulation regime, where topological transitions to high Chern number phases emerge — going beyond the traditional Haldane model. Our results not only deepen the understanding of exotic topological phases, but also offer a robust, reconfigurable, and room-temperature-compatible platform that connects quantum simulation to real-world quantum technologies.

}


\maketitle
\newpage

\section*{Introduction}

The Haldane model \cite{Haldane1988} holds profound significance in topological physics and serves as a paradigmatic model for demonstrating the capabilities of quantum simulation platforms. It shows that band topology, rather than magnetic fields or Landau levels, underpins the quantum Hall effect. This insight has catalyzed the emergence of topological phases of matter \cite{Hasan2010,Qi2011} as a vibrant research field
and inspired breakthroughs in diverse areas, including photonics \cite{Ozawa2019,Khanikaev2017}, ultracold atoms \cite{Zhu2018,Cooper2019}, and quantum information \cite{Nayak2008}.
Despite its fundamental importance, the experimental realization of the Haldane model remains challenging and has been achieved only in a few quantum systems, each representing a technical milestone in the development of the corresponding artificial quantum platform, including ultracold atoms in shaken optical lattices \cite{Jotzu2014}, graphene modulated by ultrafast light pulses \cite{Mclver2020}, moir\'e lattices in a magnetic field \cite{Zhao2024}, and superconducting quantum circuits in Floquet modulation \cite{Deng2022}. 

However, these pioneering realizations in quantum gases or solid-state materials are inherently sensitive to thermal noise and require cryogenic conditions to preserve quantum coherence. This limitation not only complicates experimental setups, but also restricts potential applications in practical devices. Moreover, thermal fragility hinders the exploration of topological phases under strong driving conditions. To break the time-reversal symmetry to achieve topologically non-trivial Bloch bands, Floquet modulation is typically introduced \cite{Rudner2020,Oka2018}, which inevitably induces heating, such that previous quantum simulation platforms have focused on topological materials with Chern numbers $C=\pm 1$ in weak driving regimes \cite{Jotzu2014,Mclver2020,Zhao2024,Deng2022}. Quantum platforms that can withstand strong driving, which leads to topological materials with higher Chern numbers, remain experimentally elusive.

Superradiance lattices (SLs) \cite{Wang2015,Chen2018} offer a versatile and highly compatible platform for simulating exotic quantum dynamics in thermal atomic ensembles under a wide range of driving conditions \cite{Cai2019,He2021,Xu2022,Mao2022}.
In this work, we demonstrate the experimental realization of the Haldane model in SLs of room-temperature atoms, where we engineer a momentum-space honeycomb SL based on an electromagnetically induced transparency (EIT) system coupled by three laser fields. By modulating the phases of the three lasers that coherently and collectively couple the atoms, we introduce next-nearest-neighbor (NNN) hopping terms attached with the required complex phase factors, effectively breaking the time-reversal symmetry. The resulting topological phases can be finely controlled by adjusting the modulation depths and phases.  The sign of the Chern number dictates the competition between two superradiant emission channels, providing an in-situ observable signature of topological phase transitions. Though the SL-based Haldane model realization was originally proposed in cold atoms \cite{WangOptica2015}, crucially, a recently developed velocity scanning tomography (VST) \cite{Wang2024} technique allows us to selectively detect the  emissions from the zero-velocity atoms such that we achieve cold-atom spectroscopic resolution in room-temperature atoms. Moreover, going beyond the conventional Haldane model with NNN hopping, our platform uniquely enables strong modulation to synthesize longer-range hopping terms. This capability unlocks the generation of complex phase diagrams featuring higher Chern numbers \cite{Sticlet2013}.

\par

\section*{Results}
\begin{figure}
    \centering
    \includegraphics[width=1\linewidth]{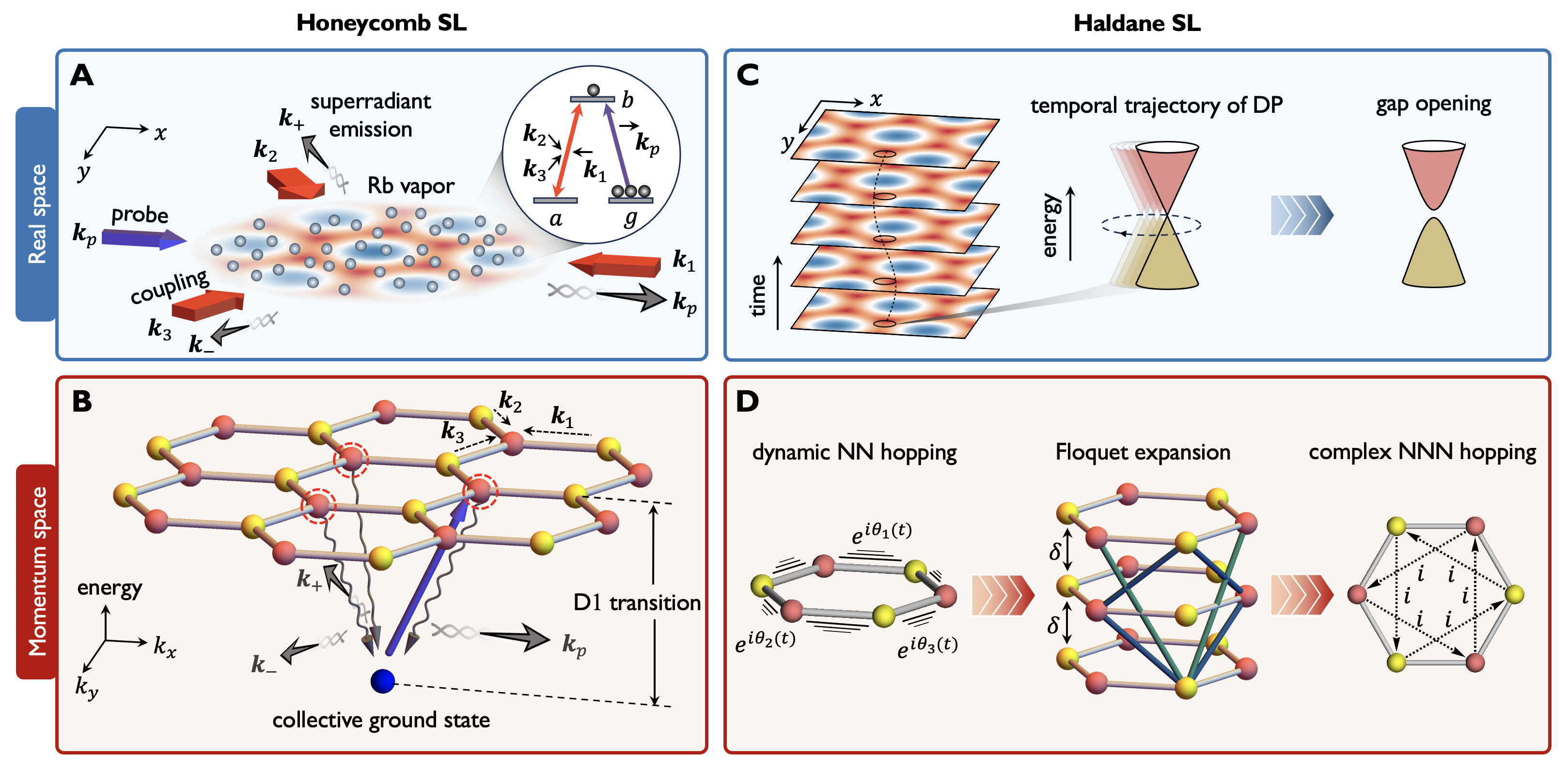}
    \caption{
    \textbf{Construction of the Haldane model in superradiance lattices, shown in both real and momentum space.} 
    \textbf{A}, Real-space representation of the honeycombe SL based on an EIT configuration (inset) with three coupling fields propagating in $\bm{k}_j$ ($j=1,2,3$) directions. A probe field along $\bm{k}_p$ excites an atomic ensemble dressed by the three coupling fields, resulting in collective emissions along $\bm{k}_\pm$. 
    \textbf{B}, Momentum-space representation of the honeycomb SL. The three-coupling-field EIT is a honeycomb tight-binding lattice in momentum space. The probe field creates a TDS $|b_{\bm{k}_p}\rangle$ from the ground state. The TDS $|b_{\bm{k}_{\pm}}\rangle$ have directional superradiant emissions in $\mathbf{k}_\pm$. 
    \textbf{C}, Time-modulated real-space energy bands. The phase modulation of the coupling fields  leads to the rotation of the DPs, breaks the time-reversal symmetry and opens bandgaps. 
    \textbf{D}, The Haldane SL in momentum space. The phase modulation of the coupling fields can be treated with a Floquet expansion, which results in an effective Hamiltonian with complex NNN hoppings.}
    \label{fig1}
\end{figure}

\begin{figure}
    \centering
    \includegraphics[width=1.0\linewidth]{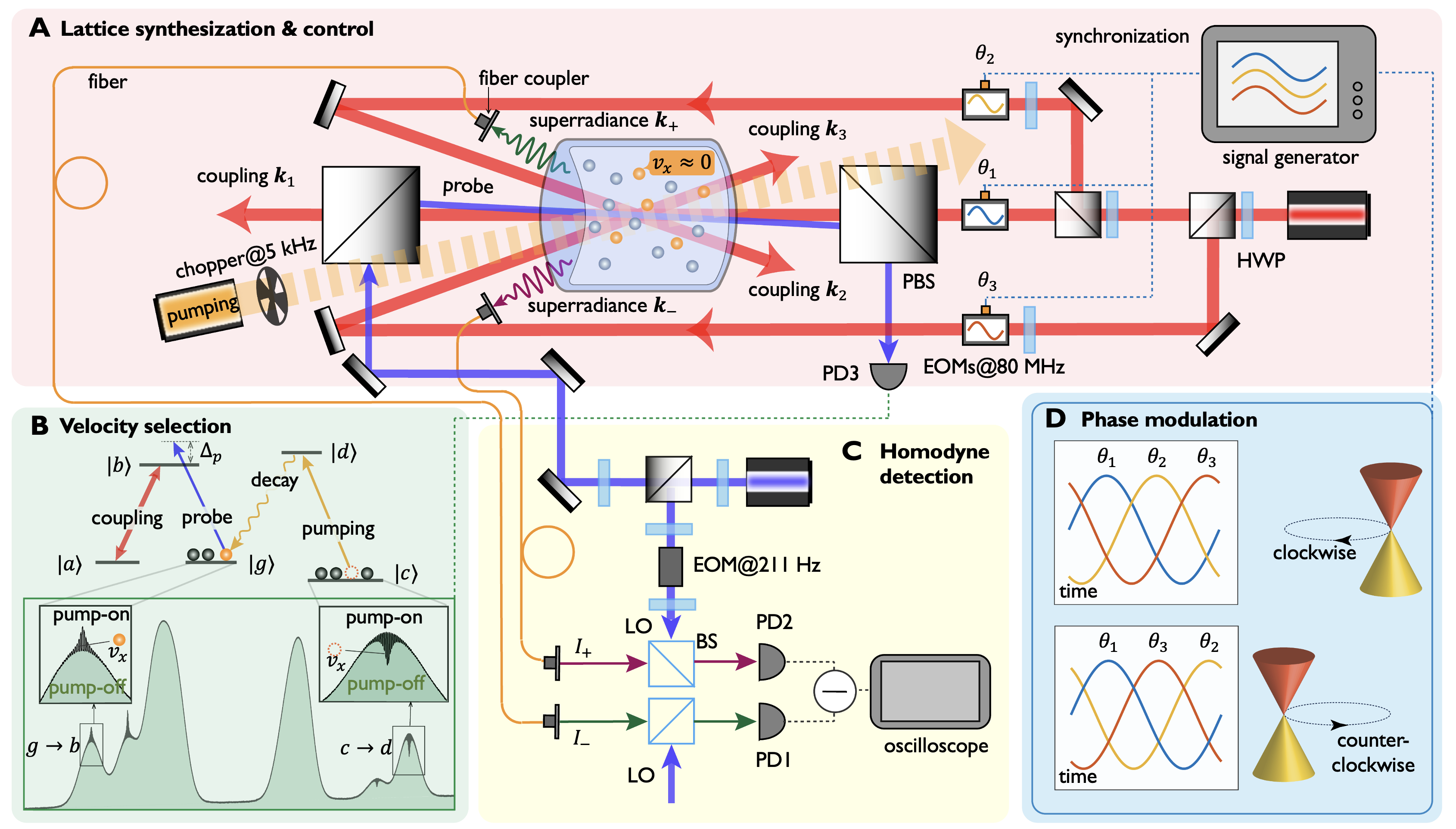}
    \caption{
    \textbf{Experimental setup for detecting the topological phase transition of Haldane superradiance lattices at room temperature.}
    \textbf{A}, The interaction between laser fields and atoms along with their phase and population modulations. Three horizontally polarized coupling fields (red beams) lie in the same plane, with $\bm{k}_1$ directed along \(-x\),  $\bm{k}_2$ and $\bm{k}_3$ oriented at angles of \(-0.5^\circ\) and \(+0.5^\circ\) to the \(x\)-axis, respectively, forming a Y-shaped configuration. The vertically polarized probe field (blue beam) intersects the plane of the coupling fields at a small angle of approximately $ 0.7^\circ$. We use polarizing beam splitters (PBSs) to combine and separate the probe and coupling fields. Three photodiodes (PDs) collect the transmission and reflections of the probe field. The atomic population is modulated by a pumping laser chopped at 5 kHz, while the phase of each coupling laser is controlled using an electro-optic modulator (EOM).
    \textbf{B}, Atomic levels and the velocity selection. The probe field couples $|g\rangle$ to the $|b\rangle$ with a detuning $\Delta_p$ while the coupling fields drive the atomic transition between $|a\rangle$ and $|b\rangle$ resonantly. A pumping laser selectively pumps the zero-velocity atoms from an idle state $|c\rangle$ to an auxiliary state $|d\rangle$, from which they subsequently decay to $|g\rangle$, thereby increasing the ground-state population. The pumping laser is chopped such that the contribution of the zero-velocity atoms can be obtained by comparing the signals when the pumping laser is on and off. The velocity selective population transfer is evidenced by the transmission spectra near the probe ($g\rightarrow b$) and pumping ($c\rightarrow d$) transitions.
    \textbf{C}, Homodyne detection of the superradiance contrast.  We mix each reflection signal with a local oscillator (LO), which is phase modulated by an EOM at frequency 211 Hz. Homodyne detection avoids interference between signals from zero-velocity atoms and those from the rest of the atomic ensemble, while also filtering out superradiant emissions originating from other Floquet replicas.
    \textbf{D}, Phase modulation of the coupling fields. The phase of the $j$th coupling field is modulated by an EOM as $\theta_j(t) = f \sin(\delta t + \phi_j)$ ($j=1,2,3$), where $\delta=80$ MHz is fixed. Assigning different modulation phases $\phi_j$ can break the time-reversal symmetry of the SLs. Here, we set the $\phi_1 = 0$ and vary $\phi_{2}$ and $\phi_{3}$ to steer the temporal trajectory of the DPs clockwisely or counter-clockwisely, thus inducing the topological phase transition.
    }
    \label{fig2}
\end{figure}

We conducted the experiment in a 2-cm-long thermal atomic vapor cell filled with naturally abundant rubidium atoms.
The SL is constructed based on the EIT configuration \cite{Fleischhauer2005}, which involves three energy levels of $^{87}$Rb atoms, a ground state $|g\rangle$, an excited state $|b\rangle$, and another ground state $|a\rangle$. Three plane-wave laser fields with wavevectors $\bm{k}_j$ ($j=1,2,3$) couple the atomic transition between $|a\rangle$ and $|b\rangle$ with spatially periodic field strengths (see Fig.~\ref{fig1}A). A weak probe field with wavevector $\bm{k}_p$ collectively excites the atoms from $|g\rangle$ to $|b\rangle$. The three coupling fields coherently modify the phase correlations of the collective excitations, resulting in two directional emissions along $\bm{k}_+ \equiv \bm{k}_p + \bm{k}_1 - \bm{k}_2$ and $\bm{k}_- \equiv \bm{k}_p + \bm{k}_1 - \bm{k}_3$. This coherent process is more intuitively understood in momentum space as a driven–dissipative honeycomb tight-binding lattice (Fig.~\ref{fig1}B). The probe field couples the ground state of $N$ atoms $|g_1,g_2,...,g_N\rangle$ to a timed-Dicke state (TDS) \cite{Scully2006,Scully2009,Araujo2016,Roof2016},
\begin{equation}
|b_{\bm{k}_p}\rangle=\frac{1}{\sqrt{N}}\sum_{m=1}^{N}e^{i\bm{k}_p\cdot\bm{r}_m}|g_1,g_2,...,b_m,...,g_N\rangle,
\label{TDS}
\end{equation}
where $\bm{r}_m$ is the position of the $m$th atom. The phase factors of TDS $|b_{\bm{k}_p}\rangle$ in Eq.~(\ref{TDS}) define its momentum $\hbar\bm{k}_p$ (we set the Planck constant $\hbar=1$ in the following), which is transferred from the probe field to the atoms. The three coupling fields couple $|b_{\bm{k}_p}\rangle$ to other TDSs accompanied by momentum transfer between the coupling fields and atoms. By collectively emitting a photon in the $i$th coupling field, the TDS $|b_{\bm{k}_p}\rangle$ is transformed to $|a_{\bm{k}_p - \bm{k}_i}\rangle$, during which the collective atomic coherence loses momentum $\bm{k}_i$. A subsequent collective absorption of a photon from the $j$th coupling field then leads to the state $|b_{\bm{k}_p - \bm{k}_i + \bm{k}_j}\rangle$. These processes collectively couple a network of TDSs, forming a honeycomb lattice in momentum space (see Fig.~\ref{fig1}B). In our specific experimental configuration, the phase correlations of the TDSs $|b_{\bm{k}_\pm}\rangle$ match those of the light \cite{He2020,Mi2021}, i.e., ${k}_\pm = k_p$, resulting in directional superradiant emissions along $\bm{k}_\pm$. These emissions are used to characterize the topological phase transitions of the SLs. The remaining TDSs have no directional emission and are therefore classified as subradiant states \cite{Scully2015}.
 
To realize the Haldane model, we periodically modulate the phases of the three coupling fields, thereby breaking the time-reversal symmetry and inducing complex NNN hopping terms. 
In the real-space Brillouin zone (BZ), which serves as the reciprocal space of the momentum-space SL, the Dirac points (DP) emerge as zero-field-strength locations resulting from destructive interference among the coupling fields, where $\sum_j e^{i\bm{k}_j \cdot \bm{r}} = 0$. We implement phase modulation on the $j$th coupling field \cite{Xu2022} as $\theta_j(t) = f \sin(\delta t + \phi_j)$, where $f$, $\delta$, and $\phi_j$ denote modulation depth, frequency, and phase, respectively. This modulation drives the DPs to follow circular trajectories \cite{Oka2009}, leading to the accumulation of a Berry phase $\pi$ for the atoms enclosed by the path in each driving period, thereby opening a bandgap (see Fig.\ref{fig1}C). The gap opening is more intuitively understood through the Floquet expansion of the SL Hamiltonian in momentum space, as shown in Fig.\ref{fig1}D. The phase modulation couples different Floquet subspaces, and from a second-order transition  we obtain NNN hopping terms with the required complex phase factors $\pm i$, realizing the Haldane SL (see the Supplementary Information (SI)).

Because momentum-space SLs lack physical edges, their topological phase transitions cannot be identified through quantum Hall responses or chiral edge currents. Instead, the transition is revealed by a sign reversal in the superradiance contrast $\eta$ when the probe field frequency is tuned to the center of the bandgap \cite{WangOptica2015},
\begin{equation}
\eta=\frac{|{c}_{\bm{k}_+}|^2-|{c}_{\bm{k}_-}|^2}{|{c}_{\bm{k}_+}|^2+|{c}_{\bm{k}_-}|^2},
\end{equation}
where $c_{\bm{k}_\pm}$ are the steady-state probability amplitudes of  $|b_{\bm{k}_\pm}\rangle$  and are directly proportional to the field strengths of their corresponding superradiant emissions. Experimentally, we determine $\eta$ by measuring the intensity contrast between the emissions from $|b_{\bm{k}_+}\rangle$ and $|b_{\bm{k}_-}\rangle$. This contrast serves as a signature of the Chern number sign, enabling us to detect and characterize topological phase transitions. 

However, Doppler broadening remains a major challenge in detecting the contrast $\eta$ contributed by zero-velocity atoms in room-temperature atomic vapor. Due to the collinear configuration of the coupling lasers (see experimental setup in Fig.~\ref{fig2}A), only the velocity in $x$-direction, $v_x$, is relevant. This velocity component effectively acts as a linear potential in the momentum-space SL \cite{Mao2022}. The thermal distribution of $v_x$ obscures the observation of the topological phase transition. To address this issue, we implement a VST technique \cite{Wang2024} combined with a homodyne detection method to isolate atoms with $v_x\approx 0$ and measure their superradiant emissions, as shown in Fig.~\ref{fig2}B and \ref{fig2}C. 

We use VST to identify the contribution of atoms with arbitrary velocities. To select zero-velocity atoms, a pumping laser is tuned to resonant with the transition between two auxiliary states $|c\rangle$ and $|d\rangle$, such that the atoms with $v_x\approx 0$ are excited and subsequently decay to the ground state $|g\rangle$, thereby slightly increasing the population of zero-velocity atoms in $|g\rangle$. The response of the zero-velocity atoms is obtained by comparing the difference between the signals when the pumping laser is on and off, controlled by an acousto-optic modulator (AOM) operating at a chopping frequency of 5 kHz. This frequency is chosen to filter out low-frequency noise while ensuring that a steady state is reached during each on/off period. When the pumping laser is on, an absorption peak appears at the center of the Doppler-broadened spectrum of the $|g\rangle \rightarrow |b\rangle$ transition, in contrast to a dip for the $|c\rangle \rightarrow |d\rangle$ transition. This indicates that the population transfer of the zero-velocity atoms from $|c\rangle$ to $|g\rangle$ (see the insets in Fig.~\ref{fig2}B). However, we cannot directly extract the superradiant emissions of the zero-velocity atoms along $\bm{k}_\pm$ by simply subtracting the superradiance intensities with the pumping laser on and off, where the result is dominated by the interference term between the superradiant emission of the zero-velocity atoms and that of all other atoms. To address this, we implement a homodyne detection module to retrieve the superradiant field amplitudes prior to performing the subtraction (see Fig.~\ref{fig2}C and details in SI).

\begin{figure}
    \centering
    \includegraphics[width=1\linewidth]{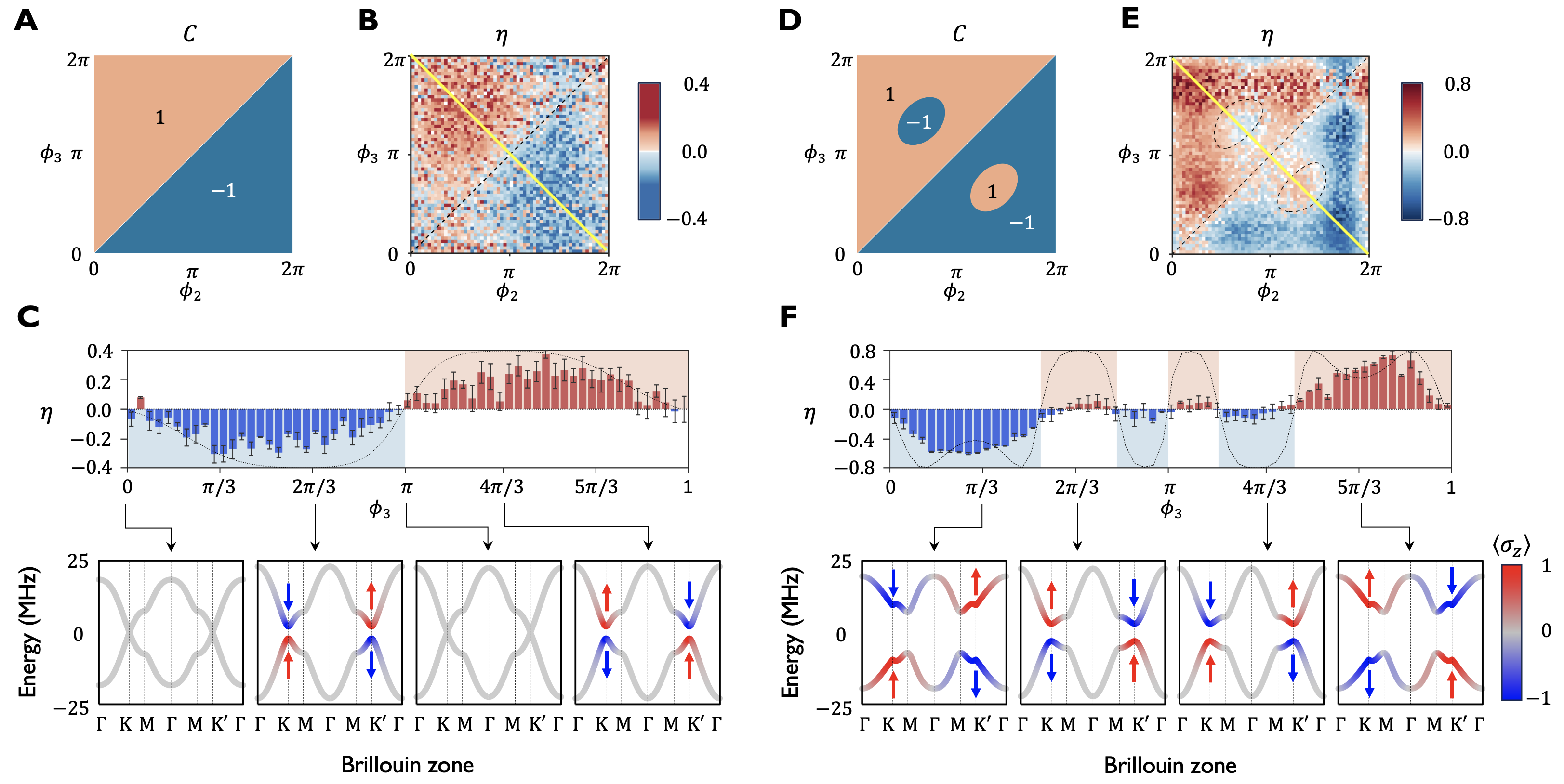}
    \caption{
        \textbf{The topological phase transition of the Haldane superradiance lattice.} $\bf{A}$, Topological phase diagram as a function of $\phi_2$ and $\phi_3$. We set $\phi_1=0$, $\Omega=10$ MHz, and $f=1.0$. The sign of the Chern number is determined by the cyclic order of $\phi_j$'s on the unit circle. $\bf{B}$, The measured data of the superradiance contrast $\eta$ with its sign indicating the topological phases in $\bf{A}$.  
        $\bf{C}$, The contrast along the yellow line in $\bf{B}$ (upper panel) and the sublattice polarization $\langle\sigma_z\rangle$ of the energy bands for $\phi_3=2\pi-\phi_2=0$, $2\pi/3$, $\pi$ and $4\pi/3$ (lower panel, from left to right). The blue and red shaded areas indicate the topological phases with $C=-1$ and $1$, respectively. The blue and red bars are the measured data of $\eta$ and the dashed line is the corresponding numerical simulation. The error bars of the experimental data are  obtained from five independent data sets. The blue and red arrows indicate the negative and positive polarizations at the DPs, corresponding to negative and positive signs of $h_z$.
        $\bf{D}$-$\bf{F}$, The same as $\bf{A}$-$\bf{C}$ but for parameters $\Omega=25$ MHz and $f=3.2$. A larger $f$ induces islands of the opposite topological phases in the original phase diagram. In the lower panel of $\bf{F}$ are the sublattice polarizations of the energy bands for $\phi_3=2\pi-\phi_2=\pi/3$, $2\pi/3$, $4\pi/3$ and $5\pi/3$ from left to right. The dashed lines in $\bf{B}$ and $\bf{E}$ highlight the phase boundary.}
    \label{fig3}
\end{figure}

Topological phases with different Chern numbers can be realized by tuning the modulation depth $f$ and the modulation phase $\phi_j$ of $\theta_j(t)$. During the modulation, the DPs undergo circular motion, with the radius set by $f$ and the rotation direction determined by the arrangement of $\phi_j$ on the unit circle (see Fig.~\ref{fig2}D). 
This Floquet modulation leads to an effective Hamiltonian $ H_{\text{eff}}(\bm{r}) = \bm{h}(\bm{r})\cdot{\bm{\sigma}}$, where $\bm{\sigma}\equiv\{\sigma_x,\sigma_y,\sigma_z\}$ is the Pauli matrix vector with $\sigma_z=|b\rangle\langle b|-|a\rangle\langle a|$. The vector $\bm{h}\equiv\{h_x,h_y,h_z\}$ characterizes the band structure of the SLs. In the case of the Haldane model with a small modulation depth $f=1.0$, the modulation induced component $h_z$ opens bandgaps at the two real-space DPs, $K$ and $K'$, where $h_x=h_y=0$. The Chern number of the lower band can be determined by $C = \{\text{sgn}[h_z(K)]\chi(K)+\text{sgn}[h_z(K')]\chi(K')\}/2$, where $\chi = \text{sgn}(\partial_{x} \bm{h}\times\partial_y\bm{h})_z$ is the chirality of the DPs. Since the chiralities of the two gapped DPs are opposite $\chi(K)=-\chi(K')=1$, the Chern number is $C=\{\text{sgn}[h_z(K)]-\text{sgn}[h_z(K')]\}/2$. For a large modulation depth $f$, additional DPs may appear and we need to count the values of $\text{sgn}(h_z) \chi $ for all the DPs to obtain the Chern number.

For a small $f=1.0$, it is sufficient to consider only the second-order transitions induced by neighboring Floquet bands, the Chern number is $C=\text{sgn}\{\sin[(\phi_1-\phi_2)/2]\sin[(\phi_2-\phi_3)/2]\sin[(\phi_3-\phi_1)/2)]\}$. In this case, the clockwise and counterclockwise rotation directions of the DPs (see Fig.~\ref{fig2}D) are associated with the Chern number $C=-1$ and $1$, respectively, as shown by the phase diagram in Fig.~\ref{fig3}A. The two topologically distinct configurations of $\phi_j$'s on the unit circle correspond to the two topological phases of the Haldane model \cite{WangOptica2015}. The phase diagram, plotted as a function of $\phi_2$ and $\phi_3$ with $\phi_1=0$, is divided into two regions by the diagonal line $\phi_2=\phi_3$. The sign of the experimentally measured superradiance contrast $\eta$ serves as an indicator of the topological phase (Fig.~\ref{fig3}B). Minor discrepancies near the diagonal corners arise from local effects \cite{WangOptica2015} when the three modulation phases are nearly identical. The topological phase transition is clearly indicated by a sign change in $\eta$ (see the upper panel in Fig.~\ref{fig3}C). In order to highlight the relation between the Chern number and the sign of $h_z$ at the DPs, we show the sublattice polarization $\langle\sigma_z\rangle$ of the eigenstates in the lower panel of Fig.~\ref{fig3}C. The opposite polarizations at the $K$ and $K'$ points indicate the opposite signs of $h_z$, which results in nonzero Chern numbers. Across the topological phase transition where the bandgap closes, $h_z$ changes its sign at the DPs.

For a large $f=3.2$, the second-order transitions between non-adjacent Floquet bands become significant, the Chern number is given by $C=\text{sgn}\{ \sum_{n=1}^\infty \sum_{j=1}^{3}J_n^2(f)\sin[n(\phi_{j+1}-\phi_j)]/n\}$, where $J_n(f)$ is the $n$th order Bessel function. Interference between different quantum paths leads to small islands of the opposite topological phase within an otherwise uniform topological phase region, similar to a ``yin and yang symbol'' (see Fig.~\ref{fig3}D). Even in this strongly modulated regime, the sign of $\eta$ remains a reliable indicator of the topological phases, with multiple topological phase transitions evidenced by sign changes in $\eta$ (see Fig.~\ref{fig3}E and~\ref{fig3}F). Across each of the transition points, the sign change of $h_z$ is accompanied by the inversion of the polarization.

\begin{figure}
    \centering
    \includegraphics[width=1\linewidth]{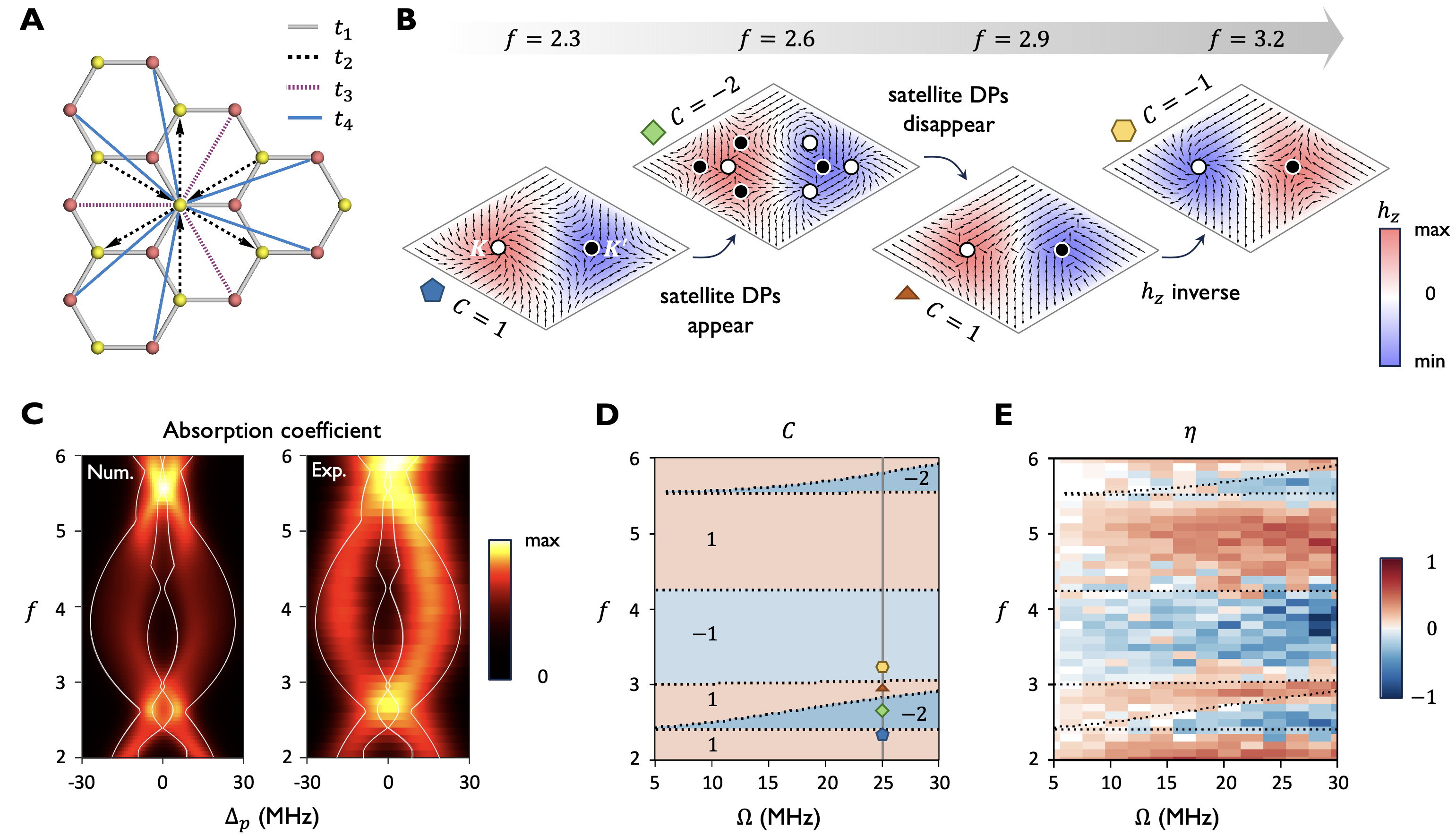}
    \caption{
        \textbf{Longer-range hoppings, satellite Dirac points and high Chern number phases.}
        $\bf{A}$, Schematic illustration of the longer-range hoppings. $t_3$ and $t_4$ are two different types of next-next-nearest-neighbor hopping terms that need to be considered. The arrows of $t_2$  denote the direction of the hoppings attached with a phase factor $i$. The other hopping strengths $t_1$, $t_3$ and $t_4$ are all real. $\bf{B}$, The DPs for different modulation strengths. The direction of $\{h_x,h_y\}$ is denoted by the in-plane arrows, while $h_z$ is denoted by the color. The modulation depths $f=2.3,2.6,2.9,3.2$ are chosen along the vertical line in \textbf{D}, marked by the corresponding polygons. For $f=2.6$, around the $K$ point ($\chi=1$, which is denoted as a white spot), three satellite DPs with opposite chirality ($\chi=-1$, which is denoted as a black spot) emerge. Similar satellite DPs emerge around the $K'$ points. The emergence of the satellite DPs leads to a topological phase transition from $C=1$ to $C=-2$.
        $\bf{C}$, The absorption spectra as functions of $f$ (along the vertical line in $\bf{D}$). The spectra represent the density of states of the energy band, with the white lines outlining the band edges. The crossings of the two lines in the middle at $\Delta_p=0$ are signatures of bandgap closing and reopening.
        $\bf{D}$, Numerically calculated topological phase diagram as a function of $\Omega$ and $f$. The polygons and vertical line at $\Omega=25$ MHz label the parameters in $\bf{B}$ and $\bf{C}$.
        $\bf{E}$, The  experimental data of the superradiance contrast $\eta$ with its sign indicating the sign of the Chern number in $\bf{C}$.  The modulation phases are fixed,  $\phi_1=0$, $\phi_2=2\pi/3$, and $\phi_3=4\pi/3$.}
    \label{fig4}
\end{figure}

Beyond the conventional Haldane model, which features only two gapped DPs in the BZ, we can tune the modulation depth $f$ to induce additional DPs, enabling topological phases with Chern numbers greater than one. In this regime, third-order transitions introduce next-next-nearest-neighbor hoppings in the honeycomb lattice (see 
$t_3$ and $t_4$ in Fig.~\ref{fig4}{A}). These longer-range hoppings lead to the formation of three satellite DPs \cite{Sticlet2013} around each original DP at the $K$ and $K'$ points, as illustrated in Fig.~\ref{fig4}{B} for $f=2.6$.
The satellite DPs are gapped and share the same chirality, which is opposite to that of the central DP, while the sign of $h_z$
remains unchanged. As a result, the Chern number changes from 1 to 
$-2$, marking a high-order topological phase transition.

The absorption spectra of the atoms, which reflect the density of states of the SLs, reveal that the emergence of satellite DPs are associated with the flattening of energy bands, evident near $f = 2.6$ and $f = 5.7$ in Fig.~\ref{fig4}C. Around these values of $f$, the nearest-neighbor hopping $t_1 \approx \Omega J_0(f)$ becomes small and comparable in magnitude to the higher-order hopping strengths $t_2$, $t_3$, and $t_4$. The band flatness arises from interference among these multiple transition pathways, leading to new valleys around the satellite DPs. Topological phase transitions occur through the closing and reopening of bandgaps, marked by the crossing of band edges, highlighted by the two white lines near $\Delta_p=0$. Although the band closing and reopening is not visible in the absorption spectra, the superradiance contrast $\eta$ accurately captures the topological phase transition in a rich phase diagram as a function of $\Omega$ and $f$ over a broad parameter space, as shown in Figs.~\ref{fig4}D and~\ref{fig4}E. The topological phase transitions between positive and negative Chern numbers are accompanied by a sign change in $\eta$. In particular, the higher-order topological phase transition from $C = 1$ to $C = -2$ is also signaled by the change in sign of $\eta$, although $\eta$ alone is insufficient to determine the precise value of the topological invariant. The exact value can instead be extracted from the winding of the Zak phase \cite{Zak1989,Xiao2010,Atala2013,Li2023} using the Wilson loop \cite{Abanin2013,Vanderbilt2018} method across the BZ, which we leave for future work. Our results demonstrate the thermal robustness and exceptional tunability of SL in the synthesis and detection of novel topological phases of the collective excitations of atoms.

\section*{Discussion}
Our results establish the SLs as a robust and versatile platform to explore topological phases at room temperature. Not only does our approach faithfully reproduce the key features of the Haldane model without requiring cryogenic conditions, but it also uniquely enables access to regimes with higher Chern numbers, thereby extending beyond the conventional Haldane paradigm. While recent work \cite{Mitra2024} has shown that the Haldane Hamiltonian can be realized at ambient temperature in monolayer hexagonal boron nitride driven by a tailored strong light pulse, the time-reversal symmetry-breaking terms in that system were too weak to realize transitions to nontrivial topological phases. Proposals for high-temperature, high-Chern-number topological phases in solid-state systems \cite{Li2025, Xie2025} have not yet been experimentally realized.

The exceptional tunability of our approach opens new avenues for investigating exotic topological phenomena that were previously inaccessible on traditional platforms. Using magnetic sublevels to encode spin degrees of freedom, we can generalize our method to simulate the quantum spin Hall state \cite{Kane2005}, which consists of two copies of the Haldane model with opposing Chern numbers. In particular, by incorporating additional coupling fields, our method can be scaled to high-dimensional band geometries and incorporated with Rydberg interactions \cite{Ding2024,Wu2024}, allowing room temperature investigation of the effect of many-body interactions in exotic phases of matter in high dimensions \cite{Yu2025}.

Moreover, operating entirely under ambient conditions positions this platform as a highly promising candidate for dynamic and reconfigurable quantum technologies. Potential applications include quantum sensors that leverage phase transitions for enhanced sensitivity and quantum routers that guide excitations based on their topological invariants. In general, our findings lay a strong foundation for advances in quantum simulation, topological photonics, and the development of functional quantum devices.

\vspace{12pt}
\setlength{\parindent}{0em}
\textbf{Code availability}

\setlength{\parindent}{2em}
The simulation code is available from the corresponding authors upon reasonable request.

\vspace{12pt}
\setlength{\parindent}{0em}
\textbf{Acknowledgements}

\setlength{\parindent}{2em}
This work is supported by the National Key
Research and Development Program of China (Grant
No. 2024YFA1408900), National Natural Science Foundation of China (Nos.~12434020, 12325412, 12404573, 12374335 and U21A20437), Zhejiang Provincial Natural
Science Foundation of China (Grants No.~LZ24A040001
and No.~R25A040015), the Fundamental Research Funds for the Central Universities (Grant No.~2024FZZX02-01-02), Innovation Program for Quantum
Science and Technology (Grant No. 2021ZD0303200), and “Pioneer” and “Leading Goose” R\&D Program of Zhejiang (Grant No. 2025C01028).

\vspace{12pt}
\setlength{\parindent}{0em}
\textbf{Author contributions}

\setlength{\parindent}{2em}
D.W.W., H.C., and X.X. conceived the project and designed the experiment. J.W., Y.L., X.L., and X.X. constructed the experimental setup, performed the measurements, and collected the data. J.D., R.M., and H.C. carried out the numerical simulations. J.W., J.D., R.M., X.X., H.C., and D.W.W. contributed to the development of the methodology. J.D., R.M., H.C., and D.W.W. conducted the theoretical analysis. All authors contributed to data analysis. H.C., X.X., and D.W.W. wrote the manuscript with feedback and input from all authors.


\begin{thebibliography} {99}

\bibitem{Haldane1988} F.~D. M. Haldane,
\newblock{Model for a quantum Hall effect without Landau levels: Condensed-matter realization of the ``parity anomaly",}
\newblock{Phys. Rev. Lett.} \textbf{61}, 2015 (1988).

\bibitem{Hasan2010} M.~Z. Hasan and C.~L. Kane,
\newblock{Colloquium: Topological insulators,}
\newblock{Rev. Mod. Phys.} \textbf{82}, 3045 (2010).

\bibitem{Qi2011} X.~L. Qi and S.~C. Zhang,
\newblock{Topological insulators and superconductors,}
\newblock{Rev. Mod. Phys.} \textbf{83}, 1057 (2011).

\bibitem{Khanikaev2017} A.~B. Khanikaev and G.~Shvets,
\newblock{Two-dimensional topological photonics,}
\newblock{Nat. Photonics} \textbf{11}, 763 (2017).

\bibitem{Ozawa2019} T.~Ozawa, H.~M. Price, A.~Amo, N.~Goldman, M.~Hafezi, L.~Lu, M.~C. Rechtsman, D.~Schuster, J.~Simon, O.~Zilberberg, and I.~Carusotto,
\newblock{Topological photonics,}
\newblock{Rev. Mod. Phys.} \textbf{91}, 015006 (2019).



\bibitem{Zhu2018} D.~W. Zhang, Y.~Q. Zhu, Y.~X. Zhao, H.~Yan, and S.~L. Zhu,
\newblock{Topological quantum matter with cold atoms,}
\newblock{Adv. in Phys.} \textbf{67}, 253 (2018).


\bibitem{Cooper2019} N.~R. Cooper, J.~Dalibard, and I.~B. Spielman,
\newblock{Topological bands for ultracold atoms,}
\newblock{Rev. Mod. Phys.} \textbf{91}, 015005 (2019).

\bibitem{Nayak2008} C.~Nayak, S.~H. Simon, A.~Stern, M.~Freedman, and S.~Das Sarma,
\newblock{Non-Abelian anyons and topological quantum computation,}
\newblock{Rev. Mod. Phys.} \textbf{80}, 1083 (2008).

\bibitem{Jotzu2014} G.~Jotzu, M.~Messer, R.~Desbuquois, M.~Lebrat, T.~Uehlinger, D.~Greif, and T.~Esslinger,
\newblock{Experimental realization of the topological Haldane model with ultracold fermions,}
\newblock{Nature} \textbf{515}, 237 (2014).

\bibitem{Mclver2020} J.~M.~Mclver, B.~Schulte, F.-U.~Stein, T.~Matsuyama, G.~Meier, and A.~Cavalleri,
\newblock{Light-induced anomalous Hall effect in graphene,}
\newblock{Nat. Phys.} \textbf{16}, 38 (2020).

\bibitem{Zhao2024} W.~Zhao, K.~Kang, Y.~Zhang, P.~Kn\"{u}ppel, Z.~Tao, L.~Li, C.~L. Tschirhart, E.~Redekop, K.~Watanabe, T.~Taniguchi, A.~F. Young, J.~Shan, and K.~F. Mak,
\newblock{Realization of the Haldane Chern insulator in a moir\'{e} lattice,}
\newblock{Nat. Phys.} \textbf{20}, 275 (2024).

\bibitem{Deng2022} J.~Deng, H.~Dong, C.~Zhang, Y.~Wu, J.~Yuan, X.~Zhu, F.~Jin, H.~Li, Z.~Wang, H.~Cai, C.~Song, H.~Wang, J.~Q. You, and D.~W. Wang,
\newblock{Observing the quantum topology of light,}
\newblock{Science} \textbf{378}, 966 (2022).

\bibitem{Rudner2020} M.~S. Rudner and N.~H. Lindner,
\newblock{Band structure engineering and non-equilibrium dynamics in Floquet topological insulators,}
\newblock{Nat. Rev. Phys.} \textbf{2}, 229 (2020).

\bibitem{Oka2018} T.~Oka and S.~Kitamura,
\newblock{Floquet engineering of quantum materials,}
\newblock{Annu. Rev. Conden. Ma. P.} \textbf{10}, 387 (2019).


\bibitem{Wang2015} D.~W. Wang, R.~B. Liu, S.~Y. Zhu, and M.~O. Scully,
\newblock{Superradiance lattice,}
\newblock{Phys. Rev. Lett.} \textbf{114}, 043602 (2015).

\bibitem{Chen2018} L.~Chen, P.~Wang, Z.~Meng, L.~Huang, H.~Cai, D.~W. Wang, S.~Y. Zhu, and J.~Zhang,
\newblock{Experimental observation of one-dimensional superradiance lattices in ultracold atoms,}
\newblock{Phys. Rev. Lett.} \textbf{120}, 193601 (2018).

\bibitem{Cai2019} H.~Cai, J.~Liu, J.~Wu, Y.~He, S.~Y. Zhu, J.~X. Zhang, and D.~W. Wang,
\newblock{Experimental observation of momentum-space chiral edge currents in room-temperature atoms,}
\newblock{Phys. Rev. Lett.} \textbf{122}, 023601 (2019).

\bibitem{He2021} Y.~He, R.~Mao, H.~Cai, J.-X.~Zhang, Y.~Li, L.~Yuan, S.-Y.~Zhu, and D.-W.~Wang,
\newblock{Flat-band localization in Creutz superradiance lattices,}
\newblock{Phys. Rev. Lett.} \textbf{126}, 103601 (2021).

\bibitem{Mao2022} R.~Mao, X.~Xu, J.~Wang, C.~Xu, G.~Qian, H.~Cai, S.~Y. Zhu, and D.~W. Wang,
\newblock{Measuring Zak phase in room-temperature atoms,}
\newblock{Light Sci. Appl.} \textbf{11}, 291 (2022).

\bibitem{Xu2022} X.~Xu, J.~Wang, J.~Dai, R.~Mao, H.~Cai, S.~Y. Zhu, and D.~W. Wang, 
\newblock{Floquet superradiance lattices in thermal atoms,}
\newblock{Phys. Rev. Lett.} \textbf{129}, 273603 (2022).

\bibitem{WangOptica2015} D.~W. Wang, H.~Cai, L.~Yuan, S.~Y. Zhu, and R.~B. Liu,
\newblock{Topological phase transitions in superradiance lattices,}
\newblock{Optica} \textbf{2}, 712 (2015).

\bibitem{Wang2024} J.~Wang, R.~Mao, X.~Xu, Y.~Lu, J.~Dai, X.~Liu, G.~Q. Liu, D.~Lu, H.~Hu, S.~Y. Zhu, H.~Cai, and D.~W. Wang,
\newblock{Velocity scanning tomography for room-temperature quantum simulation,}
\newblock{Phys. Rev. Lett.} \textbf{133}, 183403 (2024).

\bibitem{Sticlet2013} D.~Sticlet and F.~Pi\'{e}chon,
\newblock{Distant-neighbor hopping in graphene and Haldane models,}
\newblock{Phys. Rev. B} \textbf{87}, 115402 (2013).

\bibitem{Fleischhauer2005} M.~Fleischhauer, A.~Imamoglu, and J.~P. Marangos,
\newblock{Electromagnetically induced transparency: Optics in coherent media,}
\newblock{Rev. Mod. Phys.} \textbf{77}, 633 (2005).

\bibitem{Scully2006} M.~O. Scully, E.~S. Fry, C.~H. Ooi, and K.~W{\'{o}}dkiewicz,
\newblock{Directed spontaneous emission from an extended ensemble of N atoms: Timing is everything,}
\newblock{Phys. Rev. Lett.} \textbf{96}, 010501 (2006).

\bibitem{Scully2009} M.~O. Scully and A.~A. Svidzinsky,
\newblock{The super of superradiance,}
\newblock{Science} \textbf{325}, 1510 (2009).

\bibitem{Araujo2016} M.~O. Ara\'{u}jo, I.~Kre\v{s}i\'{c}, R.~Kaiser, and W.~Guerin,
\newblock{Superradiance in a large and dilute cloud of cold atoms in the linear-optics regime,}
\newblock{Phys. Rev. Lett.} \textbf{117}, 073002 (2016).

\bibitem{Roof2016} S.~J. Roof, K.~J. Kemp, and M.~D. Havey,
\newblock{Observation of single-photon superradiance and the cooperative Lamb shift in an extended sample of cold atoms,}
\newblock{Phys. Rev. Lett.} \textbf{117}, 073003 (2016).

\bibitem{He2020}  Y.~He, L.~Ji, Y.~Wang, L.~Qiu, J.~Zhao, Y.~Ma, X.~Huang, S.~Wu, and D.~E. Chang, 
\newblock{Geometric control of collective spontaneous emission,}
\newblock{Phys. Rev. Lett.} \textbf{125}, 213602 (2020).

\bibitem{Mi2021} C.~Mi, K.~S. Nawaz, L.~Chen, P.~Wang, H.~Cai, D.~W. Wang, S.~Y. Zhu, and J.~Zhang,
\newblock{Time-resolved interplay between superradiant and subradiant states in superradiance lattices of Bose-Einstein condensates,}
\newblock{Phys. Rev. A} \textbf{104}, 043326 (2021).


\bibitem{Scully2015}  M.~O. Scully, 
\newblock{Single photon subradiance: Quantum control of spontaneous emission and ultrafast readout,}
\newblock{Phys. Rev. Lett.} \textbf{115}, 243602 (2015).

\bibitem{Oka2009} T.~Oka and H.~Aoki,
\newblock{Photovoltaic Hall effect in graphene,}
\newblock{Phys. Rev. B} \textbf{79}, 081406 (2009).

\bibitem{Zak1989} J.~Zak,
\newblock{Berry’s phase for energy bands in solids,}
\newblock{Phys. Rev. Lett.} \textbf{62}, 2747 (1989).

\bibitem{Xiao2010} D.~Xiao, M.~C Chang, and Q.~Niu,
\newblock{Berry phase effects on electronic properties,}
\newblock{Rev. Mod. Phys.} \textbf{82}, 1959 (2010).

\bibitem{Atala2013} M.~Atala, M.~Aidelsburger, J.~T. Barreiro, D.~Abanin, T.~Kitagawa, E.~Demler, and I.~Bloch,
\newblock{Direct measurement of the Zak phase in topological Bloch bands,}
\newblock{Nat. Phys.} \textbf{9}, 795 (2013).

\bibitem{Li2023} G.~Li, L.~Wang, R.~Ye, Y.~Zheng, D.~W. Wang, X.~J. Liu, A.~Dutt, L.~Yuan, and X.~Chen, 
\newblock{Direct extraction of topological Zak phase with the synthetic dimension,}
\newblock{Light Sci. Appl.} \textbf{12}, 81 (2023).

\bibitem{Abanin2013} D.~A. Abanin, T.~Kitagawa, I.~Bloch, and E.~Demler,
\newblock{Interferometric approach to measuring band topology in 2D optical lattices,}
\newblock{Phys. Rev. Lett.} \textbf{110}, 165304 (2013).

\bibitem{Vanderbilt2018} D.~Vanderbilt,
\newblock{Berry phases in electronic structure theory,}
\newblock{Cambridge University Press} (2018).

\bibitem{Mitra2024} S.~Mitra, \'{A}.~Jim\'enez-Gal\'{a}n, M.~Aulich, M.~Neuhaus, R.~E. F. Silva, V.~Pervak, M.~F. Kling, and S.~Biswas,
\newblock{Light-wave-controlled Haldane model in monolayer hexagonal boron nitride,}
\newblock{Nature} \textbf{628}, 752 (2024).

\bibitem{Li2025} Z.~Li, H.~Cao, and S.~Meng,
\newblock{Light-induced above-room-temperature Chern insulators in group-IV Xenes,}
\newblock{npj Compt. Mater.}
\textbf{11}, 160 (2025).

\bibitem{Xie2025} X. Xie, K. Sun, and H. Deng,
\newblock{Polariton Chern bands in 2D photonic crystals beyond Dirac cones,}
\newblock{Phys. Rev. X}
\textbf{15}, 021061 (2025).


\bibitem{Kane2005} C.~L. Kane and E.~J. Mele,
\newblock{$Z_2$ Topological order and the quantum spin Hall effect,}
\newblock{Phys. Rev. Lett.} \textbf{95}, 146802 (2005).


\bibitem{Ding2024} D.~Ding, Z.~Bai, Z.~Liu, B.~Shi, G.~Guo, W.~Li, and C.~S. Adams,
\newblock{Ergodicity breaking from Rydberg clusters in a driven-dissipative many-body system,}
\newblock{Sci. Adv.} \textbf{10}, eadl5893 (2024).

\bibitem{Wu2024} X.~Wu, Z.~Wang, F.~Yang, R.~Gao, C.~Liang, M.~K. Tey, X.~Li, T.~Pohl, and L.~You,
\newblock{Dissipative time crystal in a strongly interacting Rydberg gas,}
\newblock{Nat. Phys.} \textbf{20}, 1389 (2024).

\bibitem{Yu2025} D.~Yu, W.~Song, L.~Wang, R.~Srikanth, S.~K.~Sridhar, T.~Chen,
C.~Huang, G.~Li, X.~Qiao, X.~Wu, Z.~Dong, Y.~He, M.~Xiao, X.~Chen, A.~Dutt, B.~Gadway, and L.~Yuan,
\newblock{Comprehensive review on developments of
synthetic dimensions,}
\newblock{Photnics Insights} \textbf{4}, R06 (2025).




\end{thebibliography}
\end{document}